\documentclass[conference]{IEEEtran}
\usepackage{verbatim}
\usepackage{longtable}
\DeclareMathAlphabet{\mathpzc}{OT1}{pzc}{m}{it}
\usepackage{fancyhdr}
\usepackage{amsmath}
\usepackage{amsfonts}
\usepackage{amssymb}
\usepackage{multirow}
\usepackage[ruled,vlined,linesnumbered]{algorithm2e}
\usepackage{stfloats}
\usepackage{etoolbox}
\usepackage{float}
\usepackage{algorithmicx}
\usepackage{fancyhdr}
\usepackage{graphicx}
\usepackage{textcomp}
\usepackage{multirow}
\usepackage{diagbox}
\usepackage{tikz}
\usepackage{enumitem}
\usepackage{url}
\usepackage{mathtools}
\usepackage{color, colortbl}
\usepackage{float}
\usepackage{subfigure}

\usepackage{cite}

\usepackage{amsmath}
\usepackage{longtable}

\usepackage{dirtytalk}
\ifCLASSOPTIONcompsoc
\usepackage[caption=false,font=normalsize,labelfont=sf,textfont=sf]{subfig}
\else
\usepackage[caption=false,font=footnotesize]{subfig}
\fi
\begin{document}
\title{\textbf{MRL-PoS:} A \textbf{M}ulti-agent \textbf{R}einforcement \textbf{L}earning based \textbf{P}roof \textbf{o}f \textbf{S}take Consensus Algorithm for Blockchain}



\author{\IEEEauthorblockN{Tariqul Islam$^{1}$, Faisal Haque Bappy$^{2}$,  Tarannum Shaila Zaman$^{3}$, \\ Md Sajidul Islam Sajid$^{4}$, and Mir Mehedi Ahsan Pritom$^{5}$}
\IEEEauthorblockA{
$^{1, 2}$ School of Information Studies (iSchool), Syracuse University, Syracuse, NY, USA\\
$ ^{3}$ Computer and Information Science, SUNY Polytechnic Institute, NY, USA\\
$ ^{4}$ Computer and Information Sciences, Towson University, Towson, MD, USA\\
$ ^{5}$ Computer Science, Tennessee Tech University, Cookeville, TN, USA\\
Email: \{mtislam@, fbappy@\}syr.edu and \{zamant@sunypoly, msajid@towson, mpritom@tntech\}.edu} 
}

\maketitle

\thispagestyle{fancy}
\lhead{This work has been accepted at the 14th Annual Computing and Communication Workshop and Conference (CCWC 2024)}
\cfoot{}

\begin{abstract}
The core of a blockchain network is its consensus algorithm. Starting with the Proof-of-Work, there have been various versions of consensus algorithms, such as Proof-of-Stake (PoS), Proof-of-Authority (PoA), and Practical Byzantine Fault Tolerance (PBFT). Each of these algorithms focuses on different aspects to ensure efficient and reliable processing of transactions. Blockchain operates in a decentralized manner where there is no central authority and the network is composed of diverse users. This openness creates the potential for malicious nodes to disrupt the network in various ways. Therefore, it is crucial to embed a mechanism within the blockchain network to constantly monitor, identify, and eliminate these malicious nodes. However, there is no \textit{one-size-fits-all} mechanism to identify all malicious nodes. Hence, the dynamic adaptability of the blockchain network is important to maintain security and reliability at all times. This paper introduces MRL-PoS, a Proof-of-Stake consensus algorithm based on multi-agent reinforcement learning. MRL-PoS employs reinforcement learning for dynamically adjusting to the behavior of all users. It incorporates a system of rewards and penalties to eliminate malicious nodes and incentivize honest ones. Additionally, MRL-PoS has the capability to learn and respond to new malicious tactics by continually training its agents.
\end{abstract}


\IEEEpeerreviewmaketitle


\section{Introduction}
Blockchain networks have evolved significantly over the last few years in many different ways. One of the core changes is the evolution of consensus algorithms. Since the introduction of Bitcoin \cite{nakamoto2008bitcoin}, Proof-of-Work has been the most popular consensus algorithm. However, due to its high computational requirements, some other consensus algorithms like Proof-of-Stake\cite{saleh2021blockchainPaper4}, PBFT\cite{castro1999practical}, and Proof-of-Authority\cite{de2018pbft} were proposed and accepted by many blockchain networks and cryptocurrencies. For example, in 2022, Ethereum completely transitioned from Proof-of-Work to Proof-of-Stake consensus, reducing overall power consumption by over 99.9\%\cite{de2023cryptocurrencies}. In terms of usefulness and efficiency, Proof-of-Stake can be a good option for private or public blockchains of any size. However, it comes with some drawbacks in terms of ensuring a fair network. In a Proof-of-Stake network, the miner with higher stakes always has the best probability to win the voting process and mine blocks, which can lead to unfair incentivization, causing other miners to lose interest in participating. Additionally, some users in the network may behave maliciously to hamper regular transaction processing or perform various types of attacks. However, malicious behavior can manifest in numerous ways, making it hard to detect and eliminate malicious nodes through a set of rules.

In this paper, we propose a novel multi-agent reinforcement learning-based Proof-of-Stake consensus algorithm named MRL-PoS. Using distributed multi-agents, our proposed approach organizes a voting process to choose the lead validator for each block. Moreover, based on the activities of the agents and their voting, penalties or rewards are assigned for each round, ultimately eliminating malicious nodes by putting them behind. The agents employ reinforcement learning to learn from user behavior and configure their parameters based on penalties or rewards to vote for the best validator each time. In this way, the agents can ensure a dynamic consensus protocol that is effective in detecting and eliminating malicious nodes from the network while also focusing on all the crucial factors important for the security and reliability of the network.

The following are the major contributions of this work:

\begin{itemize}
    \item We proposed a novel multi-agent reinforcement learning-based consensus algorithm for blockchain.
    \item We presented a penalty-reward mechanism to incentivize honest nodes and eliminate malicious nodes.
    \item We also introduced a reinforcement learning-based voting algorithm to ensure fairness in the network.
    \item We showed how MRL-PoS dynamically identifies and eliminates malicious nodes.
\end{itemize}

The remainder of the paper is structured as follows. In Section II, we presented some backgrounds about Multi-agent Reinforcement Learning and Proof-of-Stake Consensus. Then in section III, we discuss the state of existing literature. In Section IV, we present our system architecture and workflow of the MRL-PoS consensus algorithm. Section V describes how we implemented the MRL-PoS. In Section VI, we discuss some experimental scenarios related to our proposed consensus algorithm, followed by Section VII, where we discuss the future directions of this work. And finally, we conclude the paper in Section VIII.

\section{Backgrounds}
\textbf{Multi-agent Reinforcement Learning.} Multi-agent reinforcement learning (MRL) is a type of reinforcement learning in which multiple agents learn to interact with each other and their environment in order to maximize their individual rewards \cite{neto2005single}. MRL is often used in cooperative settings, where the agents are working together to achieve a common goal, but it can also be used in competitive settings, where the agents are trying to achieve their own goals at the expense of the other agents. MRL is a challenging problem because the agents need to learn to coordinate their actions in order to achieve their goals. This can be difficult, especially in large and complex environments with many agents. However, MRL has been shown to be successful in a variety of tasks, including robotics, traffic control, and resource management \cite{sartoretti2019distributed, fayaz2021transmit}, \cite{neto2005single}.\\

\textbf{Proof-of-Stake Consensus.} Proof-of-Stake (PoS) consensus protocol selects block validators based on their stake in the cryptocurrency, proportional to the total amount held. This design fosters a more energy-efficient and environmentally sustainable blockchain ecosystem \cite{saleh2021blockchainPaper4}. Validators are incentivized to act honestly, as their economic interest is directly tied to the success and stability of the network. PoS introduces a fundamentally different security model, where malicious actors risk financial penalties rather than expending computational resources. This approach not only mitigates the environmental impact associated with traditional consensus algorithms but also enhances scalability by reducing the computational overhead required for block validation \cite{chaudhry2018consensusPaper1}. Furthermore, PoS exhibits resilience to centralization, as it encourages widespread participation and discourages the concentration of power in the hands of a few mining entities\cite{vashchuk2018prosPaper5}. The emergence and adoption of PoS mark a critical step toward the evolution of blockchain technology, offering a compelling alternative for the next generation of decentralized systems.

\section{Related Works}

Blockchain consensus algorithms constitute a focal point of extensive research, and numerous comparative analyses have contributed to a nuanced understanding of this intricate field \cite{chaudhry2018consensusPaper1, bach2018comparativePaper2}. One noteworthy area of investigation within this domain is the exploration of Proof-of-Stake (PoS) mechanisms, with seminal works such as those by King et al. \cite{king2012ppcoinPaper3} and Saleh et al. \cite{saleh2021blockchainPaper4}. The latter study, conducted by Saleh et al. \cite{saleh2021blockchainPaper4}, delves deeply into the energy-efficient and secure attributes inherent in PoS, shedding light on its potential advantages over traditional consensus methods. Simultaneously, Vashchuk et al. \cite{vashchuk2018prosPaper5} critically assess distinctions between PoS and the conventional Proof-of-Work approach, providing valuable insights into the comparative strengths and weaknesses of these consensus models.

In the realm of security, dedicated protocols like Ouroboros \cite{kiayias2017ouroborosPaper7} have been developed to fortify the integrity of PoS-based systems. Additionally, privacy considerations are addressed in studies such as the one by Kerber et al. \cite{kerber2019ouroborosPaper8}, highlighting the multifaceted nature of considerations in blockchain consensus.

\begin{figure*}[t]
\centering
\includegraphics[height= 2.1in]{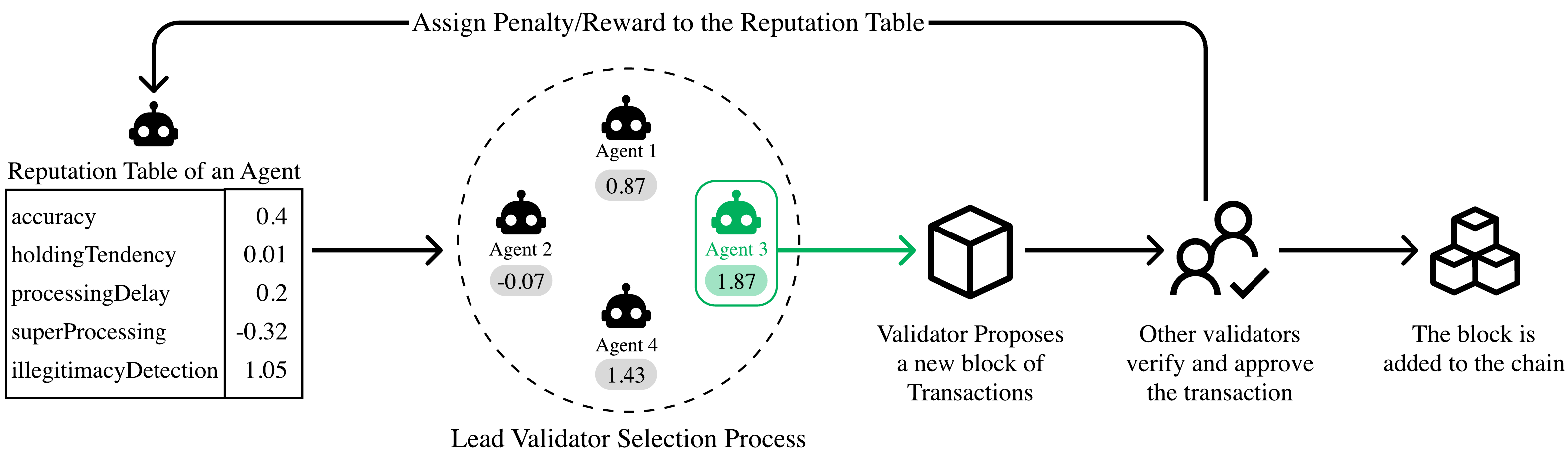}
\caption{Transaction workflow of the multi-agent consensus protocol}
\label{fig:workflow}
\end{figure*}

Beyond these foundational aspects, the discourse extends to encompass broader topics such as fairness and decentralization. Saad et al. \cite{saad2021posPaper9} contribute to this discussion, examining the implications of PoS mechanisms on fairness and the distribution of power within decentralized networks. Theoretical limitations are rigorously examined in \cite{poelstra2014distributedPaper10}, providing a critical assessment of the potential constraints that may impact the broader adoption of PoS.

Innovative perspectives for refining consensus mechanisms are proposed in various studies, including works by Bentov et al. \cite{bentov2016snowPaper11} and Li et al. \cite{li2020robustPaper12}. These contributions push the boundaries of our understanding and pave the way for the evolution of consensus protocols in blockchain systems.

While the literature extensively explores consensus algorithms, a notable gap exists in the context of dynamically detecting and eliminating malicious nodes from the network through the consensus process. Few works have ventured into this territory, and reinforcement learning has been applied in a limited number of blockchain networks to enable an intelligent consensus process \cite{zou2023optimizedPaper14, dai2019blockchainpaper15}. Importantly, none of the existing works, to date, has specifically focused on the crucial aspects of malicious node detection and dynamic adaptation of consensus mechanisms to changing user behaviors. Addressing this gap holds the potential to significantly enhance the fairness, security, and reliability of blockchain networks.

\section{System Architecture}

In this section, we present a comprehensive overview of the system architecture underpinning our proposed consensus protocol. The protocol aims to address the challenges of achieving consensus in distributed networks, ensuring robustness, efficiency, and security. Figure \ref{fig:workflow} provides a high-level depiction of the protocol's workflow, illustrating the key stages and interactions involved. The protocol initiates when a new agent joins the network, marking the commencement of the consensus process. Upon joining, a reputation table is dynamically instantiated for the newly joined agent, encompassing five core metrics that collectively define an agent's credibility within the network.

\textbf{1. Accuracy.} This metric evaluates the historical precision of an agent's validations and verifications, reflecting the agent's track record of providing accurate consensus.

\textbf{2. Transaction Holding Tendency.} The protocol considers an agent's tendency to hold transactions before validating them. A higher tendency to hold transactions might indicate that the agent could attempt to impede the flow of transactions or intentionally delay transactions with higher incentives.

\textbf{3. Processing Delay.} This metric measures an agent's efficiency in processing and validating transactions promptly. Agents with shorter processing delays contribute to the network's responsiveness.

\textbf{4. Super Processing Power.} Some agents may derive unfair advantages from relatively high computing powers, potentially gaining unexpected control over the consensus by processing the maximum number of transactions. Therefore, high computing power is a significant metric for leader selection.

\textbf{5. Ability to Detect Illegitimate Transactions.} Agents demonstrating a keen aptitude for identifying illegitimate or fraudulent transactions contribute to the network's security and integrity.

Once integrated into the network and equipped with a reputation table, the agent seamlessly transitions into the lead validator selection process. This process operates autonomously and relies on a voting mechanism, where each agent participates in voting for their peers based on the values in their respective reputation tables. Upon selection, the lead validator initiates the proposal of a new block, containing a collection of pending transactions. The proposed block undergoes verification by other validators within the network. This verification process serves a dual purpose: validating the contents of the block and reevaluating the reputation tables of participating agents. The outcome of this verification phase has implications for the agents' reputations; rewards or penalties are assigned based on the accuracy and effectiveness of their evaluations. Upon successful verification, the new block attains consensus and is committed to the main blockchain. In summary, our proposed consensus protocol integrates agent reputation, autonomous lead validator selection, transaction proposal, validation, and reputation refinement. Through these coordinated processes, the protocol endeavors to achieve consensus in a distributed network while promoting accuracy, efficiency, and security.

\section{Implementation}
Our proposed consensus mechanism necessitates the creation of a dedicated blockchain environment that has been meticulously designed to accommodate the complexities inherent in our design. This section defines the fundamental components and algorithms that make up our custom blockchain implementation, as well as the roles of agents, their decision-making processes, and the penalty-reward mechanism.

\IncMargin{1em}
\setlength{\textfloatsep}{0pt}
 \begin{algorithm}
  \caption{\textbf{\texttt{PenaltyReward()}}} 
  \SetKwFunction{BuildTree}{PenaltyReward}
  \SetKwInOut{Input}{Input}
  \SetKwInOut{Output}{Output}
  \Indm 
    \Input{Agent's action on a consensus $A_a$} 
    \Output{Updated Reputation Table of the agent $R_a$}
    \Indp
    \texttt{reputation} $\leftarrow$ 0\\

    \If{$A_a$ reached the right consensus and detected the bad actor} {
        \texttt{reputation} $\leftarrow$ +5\\
    }
    \If{$A_a$ reached the right consensus but couldn't detect the bad actor} {
        \texttt{reputation} $\leftarrow$ +2\\
    }
    \If{$A_a$ couldn't reach the right consensus but detected the bad actor} {
        \texttt{reputation} $\leftarrow$ -1\\
    }
    \If{$A_a$ couldn't reach the right consensus and couldn't detect the bad actor} {
        \texttt{reputation} $\leftarrow$ -4\\
    }
        
    \texttt{UpdateRepTab} ($R_a$, \texttt{reputation})
\end{algorithm}
\DecMargin{1em}

\textbf{Custom Blockchain Infrastructure.} The infrastructure of our blockchain environment is meticulously crafted utilizing the Go programming language. Within this tailored environment, agents emerge as autonomous entities, actively participating in the voting process. They are subject to a nuanced system of rewards and penalties that intricately shape their reputation within the network. This reputation-centric paradigm serves as the cornerstone for our dynamic consensus protocol, wherein agents iteratively adapt their behaviors in response to real-time feedback from the network. The utilization of the Go programming language ensures a robust and efficient implementation, fostering a resilient foundation for the intricate dynamics of our consensus mechanism. Algorithm 1 encapsulates the inner workings of the penalty-reward mechanism, a pivotal process that unfolds once validators complete their assessment of the various factors influencing an agent's reputation. This algorithm evaluates the outcomes of these assessments and accordingly adjusts the constant factors that drive agents' voting algorithms. By leveraging this mechanism, agents engage in a form of reinforcement learning, honing their decision-making strategies to maximize rewards and mitigate penalties. Algorithm 2 details the voting process that governs the decision-making of individual agents within the network. Agents rely on the reputation tables of their peers to inform their choices, dynamically adjusting their decisions based on the perceived reliability and competence of their counterparts. The constants $a$, $b$, $c$, $d$, and $e$ embedded within the algorithm represent pivotal factors that shape these decisions. The penalty-reward mechanism plays a pivotal role in fine-tuning these constants, allowing agents to iteratively optimize their voting strategies.




    

\IncMargin{1em}
\setlength{\textfloatsep}{0pt}
 \begin{algorithm}
  \caption{\textbf{\texttt{Voting()}}} 
  \SetKwFunction{BuildTree}{Voting}
  \SetKwInOut{Input}{Input}
  \SetKwInOut{Output}{Output}
  \Indm 
    \Input{Reputation factors of all agents $R_{all}$} 
    \Output{Votes for each agent \texttt{finalVotes}}
    \Indp


    \For{each  $R_i$ in $R_{all}$}{
        $vote \leftarrow 0$\\

        \eIf{$R_i$\texttt{.isNewAgent()}}{
            $vote \leftarrow a+b+c+d+e$\\
        }{
            $vote \leftarrow vote + a*R_i\texttt{.accuracy()}$\\
            $vote \leftarrow vote + a*R_i\texttt{.holdTendency()}$\\
            $vote \leftarrow vote + a*R_i\texttt{.processDelay()}$\\
            $vote \leftarrow vote + a*R_i\texttt{.superProcess()}$\\
            $vote \leftarrow vote + a*R_i\texttt{.illegitimate()}$\\
        }
        $finalVotes$\texttt{.append($vote$)}\\
    }
    return $finalVotes$
\end{algorithm}
\DecMargin{1em}

\begin{figure}[h]
\centering
\includegraphics[width=3.2in]{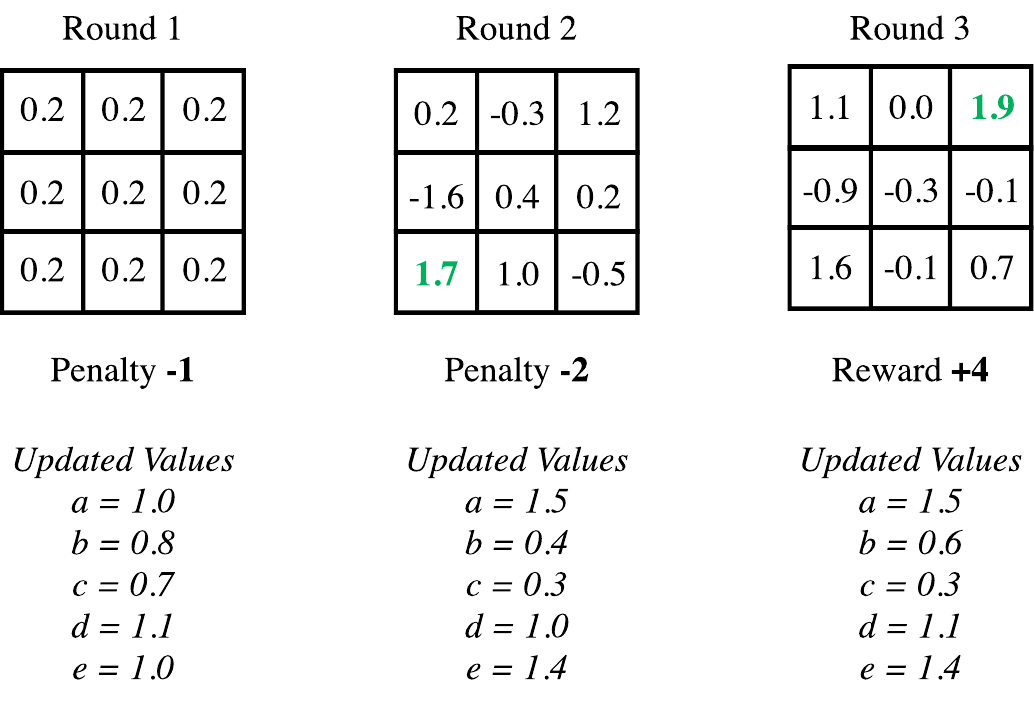}
\caption{Different Stages in Leader Selection Process}
\label{fig:voting} 
\end{figure}

Figure \ref{fig:voting} visually captures the dynamic progression of an agent's reputation through three distinct stages, as reflected by their learning table (i.e., Q-Table). In the initial stage, when all agents are newcomers to the network, the agent allocates equal votes to all peers, reflecting a uniform reputation assessment due to the absence of historical data. However, this approach leads to a penalty of -$2$, prompting the agent to recalibrate its decision-making strategy.

Upon incurring the penalty, the agent undertakes strategic adjustments to the constants governing its voting algorithm. Simultaneously, other agents accrue varying reputation values, enabling the agent to make more nuanced decisions in the subsequent stages. Nevertheless, this iteration still culminates in a penalty. In response to this continued challenge, the agent undergoes further adjustments to its constant values. Through this iterative process facilitated by the penalty-reward mechanism, the agent gradually refines its voting algorithm. The culmination of these adaptive iterations manifests in the agent accruing rewards, indicating its improved decision-making acumen and enhanced reputation within the network.

\section{Experimental Evaluation}
One of the core goals of this proposed consensus algorithm is to detect and eliminate malicious agents to ensure a secure and fair blockchain network. For this, we have formalized four experimental cases that can happen in the blockchain network.

\textbf{Case 1.} In this case, an agent can hold a few transactions while creating a new block. After winning the voting process, an agent gets to see all the pending transactions and process them to create a new block. However, if any agent holds a few transactions, it can use them to gain extra profit or launch several attacks such as Block Withholding Attack, and Sybil Attack. 

\textbf{Case 2.} The processing time for transactions can be varied highly because of the heterogeneity in computing power. But some malicious agents can delay the processing willingly hampering the flow of the whole network. By doing this, the malicious agent can open doors for attackers to exploit the network and cause desynchronization in ledgers. 

\textbf{Case 3.} In this case, an agent can have extremely high computing power (e.g., quantum computers). With this capability, it can process all the hash calculations in a very short time to get continuous rewards from the system. This will break the fairness of the system and also, the agent can use the high computing power to take over the network and perform a 51\% attack or, impersonation attack.

\textbf{Case 4.} In this case, the agent may make mistakes while processing transactions that will cause validators to throw errors and might result in delays in transaction processing. 

These aforementioned cases underscore plausible adversarial scenarios that may emerge in the event of malicious intent by a node. Consequently, to proactively detect and mitigate these threats, we have meticulously integrated five critical criteria into our reputation factors. These criteria serve as indispensable benchmarks for all participating agents in the voting process, fortifying the consensus mechanism against malicious activities and enhancing the overall resilience of the blockchain network.

\section{Discussion and Future Directions}
In this study, we introduced our preliminary ideas and findings towards creating a consensus protocol using multi-agent reinforcement learning. The aim is to establish a blockchain network that is fair, adaptable, and resistant to malicious activities. We have so far outlined the fundamental structure of the proposed consensus protocol, along with the algorithm for voting and rewards or penalties, which forms its essential logical component. In the future, our focus will be on identifying and simulating additional instances of malicious behaviors. By analyzing the outcomes of these simulations, we intend to empower the agents to recognize and counteract such malicious actions. Additionally, to enhance the equity of the reinforcement learning process, we plan to take into account a broader range of factors during the voting phase. Ultimately, our goal is to evaluate the performance of this consensus protocol within a real-world, large-scale blockchain network.

\section{Conclusion}
The primary objective of this research is to formulate and execute a distributed consensus algorithm that embodies fairness, resilience against malicious activities, and adaptability to dynamic conditions. In pursuit of this goal, we introduce an innovative consensus algorithm known as MRL-PoS. This algorithm leverages reinforcement learning to effectively address malevolent actors while incentivizing positive contributions from participants. Additionally, MRL-PoS incorporates a sophisticated voting system designed to fortify the network's security and reliability. This system evaluates various reputation factors pertinent to the candidates, ensuring a comprehensive and robust assessment. The integration of MRL-PoS into a real-world, large-scale environment is anticipated to significantly elevate the reliability and equity of blockchain networks. Furthermore, its implementation is expected to bolster defenses against malicious attacks, marking a pivotal advancement in the ongoing pursuit of secure and fair decentralized systems.

\bibliographystyle{IEEEtran}
\bibliography{main}
\end{document}